\documentclass[preprint,showpacs,preprintnumbers,amsmath,amssymb]{revtex4}

\usepackage{graphicx}
\usepackage{dcolumn}
\usepackage{bm}

\begin{document}

\title{Localization of Two-Dimensional Quantum Walks}

\author{Norio Inui, Yoshinao Konishi}
\email{inui@mie.eng.himeji-tech.ac.jp, tm02m018@mie.eng.himeji-tech.ac.jp}
\affiliation{%
Graduate School of Engineering, Himeji Institute of Technology, \\
2167, Shosha, Himeji, Hyogo, 671-2201, Japan\\}

\author{Norio Konno}
\email{norio@mathlab.sci.ynu.ac.jp}
\affiliation{%
Department of Applied Mathematics, 
Yokohama National University, 
79-5 Tokiwadai, Yokohama, 240-8501, Japan\\}

\date{\today}

\begin{abstract}
The Grover walk, which is related to the Grover's search algorithm on a quantum computer, 
is one of the typical discrete time quantum walks. However, a localization of the two-dimensional 
Grover walk starting from a fixed point is striking  different from other types of quantum walks. 
The present paper explains the reason why the walker who moves according to the degree-four Grover's
 operator can remain at the starting point with a high probability. 
It is shown that the key factor for the localization is due to the degeneration of eigenvalues of 
the time evolution operator. In fact, the global time evolution of the quantum walk on a large lattice 
is mainly determined by the degree of degeneration. The dependence of the localization on the initial
 state is also considered by calculating the wave function analytically.
\end{abstract}

\maketitle

\section{\label{sec:S1} Introduction}

The quantum walks are roughly classified into discrete time quantum walks 
\cite{Aharonov1993,Mayer1996,Nayak2000,Childs2002a,Brun2003A,Brun2003B} and
continuous time quantum walks \cite{Farhi1998,Aharonov1998}. 
We focus on the discrete time quantum walks on a square lattice. 
The study of the discrete time quantum walks was begun by Aharonov et al. \cite{Aharonov1993} 
in the early 1990s,  then it has been investigated by a number of groups. 
The discrete time quantum walk evolves by repeating simple quantum operations, 
and it is expected to be realized in a quantum computer. 
The Grover's search algorithm \cite{Grover1997}, 
which is one of the most famous quantum algorithms,
 is especially related to a discrete quantum walk \cite{Childs2002b,Childs2003}. 
Recently Shenvi et al. \cite{Shenvi2002} actually proved that a discrete, coined quantum walk 
can equal Grover's algorithm. For an introduction of the implementation by a quantum computer,
 see Travaglione and Milburn \cite{Travaglione2002}, for example.

The recent concentrated studies make clear mathematical properties of the one-dimensional 
quantum walks. In particular the one-dimensional Hadamard walk is  studied in detail
 \cite{konno2002a,konno2002b,konno2002c,konno2002d,Bednarska2003,Inui2003}.
In contrast with one-dimensional quantum walks, little about high dimensional
quantum walks is known \cite{Mackay2002,Moore2002,Kempe2003,Tregenna2003,Grimmett2003}.
 Thus the purpose of this study is to investigate a two-dimensional quantum walk called ``Grover walk". 
 A pioneering work for the Grover walk was done by Mackay et al. \cite{Mackay2002}. 
Very recently Tregenna et al. \cite{Tregenna2003} showed numerically that the quantum walker
 who is controlled by the Grover's operator is observed at an initial location with a high probability. 
In this paper, this phenomenon is referred to as ``localization".  They showed  also  that the quantum walker
 starting from a special initial state spreads out numerically.

The first question we have to ask here is whether the localization remains even after a sufficiently large time.
 Unfortunately numerical simulations can not give us the exact answer on this problem.
 Hence it follows that we have to calculate the wave function rigorously. 
Secondly we ask why the localization is observed only in the Grover walk. 
There are many different quantum walks, however, the localization is not observed except 
the Grover walk in our knowledge. Thirdly we would like to know the dependence of the localization 
on the initial state.
We will answer these questions in following sections. 

The paper is organized as follows. After defining the Grover walk in section $\mbox{I\hspace{-0.5mm}I}$, 
we calculate eigenvalues and eigenvectors of the time evolution operator to obtain 
the wave function in section I\hspace{-0.5mm}I\hspace{-0.5mm}I. 
Section I\hspace{-0.5mm}V treats the wave function at the origin and the time-averaged probability. Using the results we show that localization remains even if the system size is infinity. In section V, we concentrate our attention to the localization on an infinite lattice and explain the reason why the Grover walk is special. Furthermore we consider the dependence of the 
localization on the initial sate and show that the localization disappears at a certain initial state.

\section{\label{sec:S2} Definition of the two-dimensional quantum walks}

\subsection{Time evolution of the two-dimensional quantum walks}

The Grover walk considered here is a kind of discrete time quantum walks. 
Thus we begin with defining the two-dimensional quantum walk
on the square lattice $Z_{N}=\{(x,y) \in Z^{2} \,|\, -(N-1)/2 \leq x \leq (N-1)/2,\,  
-(N-1)/2 \leq y \leq (N-1)/2\}$
with periodic boundary condition. In this paper we assume that the system size $N$ is odd.
There are four quantum states at each  site:``R",``L",``U" and ``D"
corresponding to right, left, up and down, respectively. The value of wave function for one of these states $S \in \{R, L, U, D \}$ at the position $(x,y)$ and time $t$ is written by
$|S,x,y,t \rangle$. The time evolution of $|S,x,y,t \rangle$ is determined as follows:
\begin{eqnarray}
|R,x,y,t+1\rangle = 
a_{11}|R,x-1,y,t\rangle+a_{12}|L,x-1,y,t\rangle+a_{13}|U,x-1,y,t\rangle+a_{14}|D,x-1,y,t\rangle, \nonumber \\
|L,x,y,t+1\rangle =
a_{21}|R,x+1,y,t\rangle+a_{22}|L,x+1,y,t\rangle+a_{23}|U,x+1,y,t\rangle+a_{24}|D,x+1,y,t\rangle,  \nonumber \\
|U,x,y,t+1\rangle =
a_{31}|R,x,y-1,t\rangle+a_{32}|L,x,y-1,t\rangle+a_{33}|U,x,y-1,t\rangle+a_{34}|D,x,y-1,t\rangle,  \nonumber \\
|D,x,y,t+1\rangle =
a_{41}|R,x,y+1,t\rangle+a_{42}|L,x,y+1,t\rangle+a_{43}|U,x,y+1,t\rangle+a_{44}|D,x,y+1,t\rangle.\,  
\label{eqn:evolution}
\end{eqnarray}
This evolution is characterized by the next matrix: 
\begin{eqnarray}
A=
\left[
\begin{array}{cccc}
 a_{11} & a_{12} & a_{13} &  a_{14}\\
 a_{21} & a_{22} & a_{23} &  a_{24}\\
 a_{31} & a_{32} & a_{33} &  a_{34}\\
 a_{41} & a_{42} & a_{43} &  a_{44}
\end{array}
\right].
\label{eqn:matA}
\end{eqnarray}
The matrix corresponding to the Grover walk is defined by
\begin{eqnarray}
A_{0}=
\left[
\begin{array}{rrrr}
 -\frac{1}{2} & \frac{1}{2}  & \frac{1}{2}  &  \frac{1}{2} \\
  \frac{1}{2} & -\frac{1}{2} & \frac{1}{2}  &  \frac{1}{2} \\
  \frac{1}{2} & \frac{1}{2}  & -\frac{1}{2} &  \frac{1}{2} \\
  \frac{1}{2} & \frac{1}{2}  & \frac{1}{2} &  -\frac{1}{2}
\end{array}
\right].
\label{eqn:matA0}
\end{eqnarray}
We  introduce here other two-dimensional quantum walks to compare 
with the Grover walk by setting following matrices:
\begin{eqnarray}
A_{1} &=&
\left[
\begin{array}{rrrr}
 0                  &  0                  & -\frac{1}{\sqrt{2}}  & \hspace{3mm}\frac{1}{\sqrt{2}} \\
 0                  &  0                  &  \frac{1}{\sqrt{2}}  & \frac{1}{\sqrt{2}}  \\
 \hspace{2mm}\frac{1}{\sqrt{2}} & -\frac{1}{\sqrt{2}} &  0                   &      0  \\
 \frac{1}{\sqrt{2}} & \frac{1}{\sqrt{2}}  &  0                   &      0
\end{array}
\right],  \label{eqn:matA1} \\
A_{2} &=&
\left[
\begin{array}{cccc}
-\frac{1}{\sqrt{3}}  &  0                   &  \frac{1}{\sqrt{3}}  &  \hspace{2mm}\frac{1}{\sqrt{3}} \\
 0                   & -\frac{1}{\sqrt{3}}  & -\frac{1}{\sqrt{3}}  &  \frac{1}{\sqrt{3}}  \\
\frac{1}{\sqrt{3}}   & -\frac{1}{\sqrt{3}}  & \frac{1}{\sqrt{3}}   &  0  \\
 \frac{1}{\sqrt{3}}  & \frac{1}{\sqrt{3}}   &  0                    &  \frac{1}{\sqrt{3}}
\end{array}
\right]. 
\label{eqn:matA2}
\end{eqnarray}

We define the wave function of the total state at time $t$ by
\begin{eqnarray}
\psi(t)&=&(\psi_{0,0}(t),\psi_{1,0}(t),\psi_{2,0}(t),\ldots, \psi_{N-1,0}(t), \\ 
       & & \hspace{3mm} \psi_{0,1}(t),\psi_{1,1}(t),\psi_{2,1}(t),\ldots,
\psi_{x,y}(t), \ldots, \psi_{N-1,N-1}(t))^{T},
\label{eqn:Psi1}
\end{eqnarray}
where $T$ means the transposed operator and 
\begin{eqnarray}
\psi_{x,y}(t)=(|R,x,y,t\rangle, |L,x,y,t\rangle ,|U,x,y,t\rangle, |D,x,y,t\rangle)^{T}.
\label{eqn:Psi2}
\end{eqnarray}
If the initial state $\psi(0)$ is given, then the wave function $\psi(t)$ is calculated by the iteration (\ref{eqn:evolution}). The iteration can be expressed  more compactly  by introducing a $4N \times 4N$ unitary matrix $M$ satisfying $\psi(t+1)=M \psi(t)$.

The probability of 
observing the quantum walker  at a given point $(x,y)$ and
time $t$ starting from an initial state $\psi(0)$ is defined by
\begin{eqnarray}
P(x,y,t;\psi(0)) =\sum_{S \in \{ R, L, U, D \}}\langle S, x,y,t|S,x,y,t\rangle.
\label{eqn:defProb}
\end{eqnarray}
Fig. 1 shows the probability distribution $P \equiv P(x,y,t;\psi(0))$ 
corresponding to matrices (a) $A_{0}$ (Grover walk), (b) $A_{1}$ and (c) $A_{2}$ at $t=31$ on the lattice with $N=51$ starting from a pure initial
state $\psi(0)= (1,0,\ldots ,0)^{T}$. In contrast with (b) and (c) cases, a localization at the origin can be seen only the Grover walk case (a).

\section{\label{sec:S3} Eignenvalues and eigenvectors of the matrix $M$}

\subsection{Eigenvalues}

To express the wave function as a function of $t$ explicitly we consider
the eigenvalues and eigenvectors of the matrix $M$. These are easily obtained by the using the Fourier transform.
According to the previous studies \cite{Aharonov1998}, the eigenvaleus of the matrix  $M$
are given by a set of eigenvalues of the following matrix
\begin{eqnarray}
H_{n,m}(A)=
\left[
\begin{array}{cccc}
\omega^{-n} &     0        &     0         & 0 \\
0           &  \omega^{n}  &     0         & 0 \\
0           &     0        &  \omega^{-m}  & 0 \\
0           &     0        &     0         &  \omega^{m} 
\end{array}
\right]
\left[
\begin{array}{cccc}
 a_{11} & a_{12} & a_{13} &  a_{14}\\
 a_{21} & a_{22} & a_{23} &  a_{24}\\
 a_{31} & a_{32} & a_{33} &  a_{34}\\
 a_{41} & a_{42} & a_{43} &  a_{44}
\end{array}
\right],
\label{H}
\end{eqnarray}
where $\omega=e^{2 \pi i/N}$. 
The integers $n$ and $m$ are quantum numbers in a wave number space and
they take values between 0 and $N-1$.
Since there are four components in $H_{n,m,}(A)$,
the number of eigenvalues is $(4N)^{2}$, 
if not consider the degeneration of eigenvalues, and each eigenvalue is labeled by $n,m \in \{0,1, \ldots, N-1 \}$ and $k \in \{1,2,3,4 \}$. As a result, when $n \neq m$, the eigenvalues of the matrix $M$ corresponding to the Grover walker, $\lambda_{n,m,k}$, are given by
\begin{eqnarray}
\lambda_{n,m,1} &=& -1, \\
\lambda_{n,m,2} &=&  1, \\
\lambda_{n,m,3} &=&
    \frac{-\cos \xi_{m} - \cos \xi_{n} - 
         {\sqrt{-4 + {\left( \cos \xi_{m} + \cos \xi_{n} \right) }^2}}}{2}, \\
\lambda_{n,m,4} &=&
    \frac{-\cos \xi_{m} - \cos \xi_{n} + 
         {\sqrt{-4 + {\left( \cos \xi_{m} + \cos \xi_{n} \right) }^2}}}{2},
\label{eqn:lambda}
\end{eqnarray}
where $\xi_{j}=2 j \pi/N$.
When $n = m$, the eigenvalues are written as
\begin{eqnarray}
\lambda_{n,n,1} &=& -1, \\
\lambda_{n,n,2} &=&  1, \\
\lambda_{n,n,3} &=& -\omega^{n}, \\
\lambda_{n,n,4} &=& -\omega^{-n}.
\label{eqn:lambda2}
\end{eqnarray}

\subsection{Eigenvectors}

We write the eigenvectors corresponding to the eigenvalues $\lambda_{n,m,k}$ as  $\phi_{n,m,k}$
 and we let $\phi_{i,n,m,k}$ be  $i$-th element  of  $\phi_{n,m,k}$. 
If $-(N-1)/2 \leq x,y \leq (N-1)/2$ and $1 \leq j \leq 4$,  
we find integers $x,y$ and $j$ satisfying an equation $4 N y+4 x+j+2N^{2}-2=i$ for  a given natural
number $i$. Using these $x,y$ and $j$, the $i$-th element  of $\phi_{n,m,k}$, 
which generates an orthonormal basis is given by
\begin{eqnarray}
\phi_{i,n,m,k} &=& \frac{v_{j,n,m,k} \omega^{n x+m y}}{N
\sqrt{\sum_{j=1}^{4} |v_{j,n,m,k}|^{2}}},
\end{eqnarray} 
where $v_{j,n,m,k}$ is the $j$-th element of the eigenvector of $H_{n,m,k}$.
We show a set of eigenvectors of $v_{n,m,k} \equiv (v_{1,n,m,k},\cdots,v_{4,n,m,k})^{T}$ in the following:   

Case 1: $m=0, n>0, k=1$

\begin{eqnarray}
v_{n,0,1}=
\left[
\begin{array}{c}
1  \\
-1 \\ 
0  \\
0
\end{array}
\right],
\label{v1}
\end{eqnarray}

Case 2: $m=0,n>0$ and $k>1$ 

\begin{eqnarray}
v_{n,0,k}=
\left[
\begin{array}{c}
\lambda_{n,0,k}+\beta_{n}   \\

\lambda_{n,0,k}+\beta_{n}   \\
\beta_{n}\lambda_{n,0,k}+\beta_{n} \\
2\lambda_{n,0,k}^{2}+\beta_{n}\lambda_{n,0,k}-\beta_{n}
\end{array}
\right],
\label{v2}
\end{eqnarray}

Case 3: $n=m$

\begin{eqnarray}
\{ v_{n,n,k}|1 \leq k \leq 4 \}=
\left \{
\left[
\begin{array}{c}
-\alpha_{n}  \\
1 \\ 
-\alpha_{n}   \\
1
\end{array}
\right],
\hspace{5mm}
\left[
\begin{array}{c}
\alpha_{n}  \\
1 \\ 
\alpha_{n}   \\
1
\end{array}
\right],
\hspace{5mm}
\left[
\begin{array}{c}
0  \\
-1 \\ 
0   \\
1
\end{array}
\right],
\hspace{5mm}
\left[
\begin{array}{c}
-1  \\
0 \\ 
1   \\
0
\end{array}
\right]
\right\},
\label{v3}
\end{eqnarray}

Case 4: $n+m=N$ 

\begin{eqnarray}
\{ v_{n,N-n,k} | 1 \leq k \leq 4 \}=
\left \{
\left[
\begin{array}{c}
1  \\
-1/\alpha_{n} \\ 
-1/\alpha_{n}   \\
1
\end{array}
\right],
\hspace{5mm}
\left[
\begin{array}{c}
1  \\
1/\alpha_{n} \\ 
1/\alpha_{n}   \\
1
\end{array}
\right],
\hspace{5mm}
\left[
\begin{array}{c}
0  \\
-1   \\ 
1   \\
0
\end{array}
\right],
\hspace{5mm}
\left[
\begin{array}{c}
-1  \\
0   \\ 
0   \\
1
\end{array}
\right]
\right\},
\label{v4}
\end{eqnarray}

Case 5: otherwise
\begin{eqnarray}
v_{n,m,k}=
\left[
\begin{array}{c}
\alpha_{n}^{2} \lambda^{2}_{n,m,k}
+(\alpha_{n} +\alpha_{n}^{2} \beta_{m}) \lambda_{n,m,k}
+\alpha_{n}\beta_{m} \\
\lambda_{n,m,k}^{2}
+(\alpha_{n}+\beta_{m}) \lambda_{n,m,k} 
+\alpha_{n}\beta_{m} \\ 
\alpha_{n} \beta_{m} \lambda_{n,m,k}^{2}
+(\beta_{m}+\alpha_{n}^{2} \beta_{m}) \lambda_{n,m,k} 
+\alpha_{n}\beta_{m}\\
2\alpha_{n} \lambda_{n,m,k}^{3}
+(1+\alpha_{n}^{2}+\alpha_{n}\beta_{m})\lambda_{n,m,k}^{2}
-\alpha_{n}\beta_{m}
\end{array}
\right],
\label{v5}
\end{eqnarray}
where $\alpha_{n}=\omega^{-n}$ and $\beta_{m}=\omega^{-m}$.

\section{\label{sec:S4}Wave function of the Grover walk}

\subsection{Expansion of wave function in terms of eigenvalues}

Since we have obtained the complete eigenvalues and normalized orthogonal 
 eigenvectors, we can express the wave function as a function of
time. Before we present the wave function, we number the states R, L, U, and D from 1 to 4, respectively.
Let $l(S)$ be the number of the state ``S". Then we have the wave function $|S,x,y,t \rangle$:
\begin{eqnarray}
|S,x,y,t \rangle &=& \sum_{j=1}^{4N^{2}} \sum_{n=0}^{N-1} \sum_{m=0}^{N-1} \sum_{k=1}^{4} 
\phi_{i,n,m,k}\phi_{j,n,m,k}^{\ast}  \psi_{j}(0) \lambda_{n,m,k}^{t},
\label{eqn:wf}
\end{eqnarray}
where $i=4 N y+4 x+l(S)+2N^{2}-2$ and $\psi_{j}(0)$ is the $j$-th element of the initial vector $\psi(0)$.

As shown in (\ref{eqn:lambda}), the eigenvalues degenerate strongly. Thus we try to expand the
wave function by distinct eigenvalues. The eigenvalue $\lambda_{n,m,1}=-1$  always exists for any combination
$n$ and $m$. Furthermore the  $\lambda_{n,m,3}$ and $\lambda_{n,m,4}$ become $-1$ for $n=m=0$. If $n>0$ and $m>0$, 
then the eigenvalues $\lambda_{n,m,k}$ are distinct for fixed $n$ and $m$. Therefore
the condition $\lambda_{n,m,k}=\lambda_{n',m',k}$ is equivalent to the following condition
\begin{eqnarray}
\cos \xi_{m} + \cos \xi_{n}=\cos \xi_{m'} + \cos \xi_{n'}.
\label{eq:degcond}
\end{eqnarray}
A set of a pair $(n', m')$ belonging to the same eigenvalue $\lambda_{n,m,k}$ for $k>2$ and  $n>0$ is given by
\begin{eqnarray}
\Omega(n,m)=
\left \{
\begin{array}{l}
\{ (n,0), (0,n), (N-n,0), (0,N-n) \} \hspace{3mm} \mbox{for} \hspace{3mm} m=0, \\
\{ (n,n), (n,N-n) \} \hspace{3mm} \mbox{for} \hspace{3mm} n=m,  \\
\{ (n,m), (n,N-m), (N-n,m),(N-n,N-m) \nonumber, \\
\,\, (m,n),(m,N-n), (N-m,n), (N-m, N-n)\} \hspace{3mm} \mbox{otherwise}. 
\end{array}
\right.
\label{eqn:sets}
\end{eqnarray}
To express the wave function compactly using the set $\Omega(n,m)$ we define the following functions:
\begin{eqnarray}
c_{i,j,0,0,k} &=& 
\frac{v_{i,0,0,k}v_{j,0,0,k}^{\ast} \psi_{j}(0)}
{\sqrt{\sum_{i=1}^{4} |v_{i,0,0,k}|^{2}}\sqrt{\sum_{j=1}^{4} |v_{j,0,0,k}|^{2}}}, \label{eqn:c0}
\\
c_{i,j,n,m,k} &=& \sum_{n',m' \in \Omega(n,m)} 
\frac{v_{i,n',m',k}v_{j,n',m',k}^{\ast} \psi_{j}(0)}
{\sqrt{\sum_{i=1}^{4} |v_{i,n',m',k}|^{2}}\sqrt{\sum_{j=1}^{4} |v_{j',n',m',k}|^{2}}} \nonumber \\
& & \hspace{60mm} \mbox{for} \hspace{2mm} n>0  \hspace{2mm} \mbox{and} \hspace{2mm} m>0.
\label{eqn:c1}
\end{eqnarray}
We note here the reason why the system size is restricted to odd in this paper. In the case of odd, the degree of
degeneration of eigenvalues is eight at the most except for the eigenvalues $-1$ and $1$. On the other, the eigenvalues
with large degree of degeneration exist in addition to the $1$ and $-1$ in the case of even. 
This fact does not lead us to fatal 
difficulty, but the calculation becomes more complicated  than that in the odd case.

We now have another expression of wave function:
\begin{eqnarray}
|S,x,y,t \rangle & =& \frac{1}{N^{2}}\sum_{j=1}^{4N^{2}} 
\left[ C_{i,j,1}+C_{i,j,-1} (-1)^{t} 
+
\sum_{n=1}^{\frac{N-1}{2}}
\sum_{k=3}^{4} c_{i,j,n,0,k}\lambda_{n,0,k}^{t} \right. \nonumber \\
& &
\hspace{15mm}
+
\left.
\sum_{n=1}^{N-1}
\sum_{k=3}^{4} c_{i,j,n,n,k}\lambda_{n,n,k}^{t}
+ 
\sum_{n=1}^{\frac{N-3}{2}} 
\sum_{m=n+1}^{\frac{N-1}{2}}
\sum_{k=3}^{4} c_{i,j,n,m,k}\lambda_{n,m,k}^{t}
\right], 
\label{eqn:wf2}
\end{eqnarray}
where 
\begin{eqnarray}
C_{i,j,1} &=&c_{i,j,0,0,2}
+\sum_{n=1}^{\frac{N-1}{2}}c_{i,j,n,0,2}
+\sum_{n=1}^{N-1}c_{i,j,n,n,2}
+
\sum_{n=1}^{\frac{N-3}{2}} 
\sum_{m=n+1}^{\frac{N-1}{2}} c_{i,j,n,m,2},  \label{eqn:Cp} \\
C_{i,j,-1} &=&
c_{i,j,0,0,1}+c_{i,j,0,0,3}+c_{i,j,0,0,4}
+\sum_{n=1}^{\frac{N-1}{2}} c_{i,j,n,0,1}+
\sum_{n=1}^{N-1} c_{i,j,n,n,1}
+ \sum_{n=1}^{\frac{N-3}{2}} 
\sum_{m=n+1}^{\frac{N-1}{2}} c_{i,j,n,m,1}.
\label{eqn:Cm}
\end{eqnarray}
and $i=4 N y+4 x+l(S)+2N^{2}-2$. 
In the formula (\ref{eqn:wf2}), $C_{i,j,1}$ and $C_{i,j,-1}(-1)^{t}$ give the contribution of
the eigenvalue $1$ and $-1$ to the wave function.
The remaining terms are corresponding to the eigenvalues for $k=3$ and 4.

We move on more specific cases. In order to show the localization we calculate $|R,0,0,t \rangle$
for the Grover walk starting from a pure state $|R,0,0,0 \rangle=1$.  We refer this special initial state
as $\phi_{R}$. The values
in (\ref{eqn:c0}), (\ref{eqn:c1}) and (\ref{eqn:Cp}), (\ref{eqn:Cm}) can be obtained after some algebra by
\begin{eqnarray}
&&C_{1,1,1}=\frac{N^{2}}{4} ,\hspace{3mm} C_{i,1,-1}=\frac{1}{2}+\frac{N^{2}}{4}, 
\label{eqn:C1}
\\
&&c_{1,1,n,0,k}=1,  \hspace{3mm} \mbox{for} \hspace{3mm} \forall k, \\
&&c_{1,1,n,n,1}=c_{1,1,n,n,2}=\frac{1}{2},\hspace{2mm}c_{1,1,n,n,3}=0, \hspace{2mm}c_{1,0,n,n,4}=1, \\
&&c_{1,1,n,m,k}=2,  \hspace{3mm} \mbox{for} \hspace{3mm} \forall k. 
\label{eqn:Rc}
\end{eqnarray}

\section{Time-averaged probability}

\subsection{Definition of time-averaged probability}

We begin with considering the probability $P(S,t;\phi_{0},N)$ that a walker in the state ``S" 
at the origin on a square lattice with size $N$ starting 
form an  initial state $\phi_{0}$. The probability  $P(S,t;\phi_{0},N)$ is calculated from the relation
\begin{eqnarray}
P(S,t;\phi_{0},N) \equiv \langle S, 0,0,t | S,0,0,t \rangle.
\label{eqn:Ps}
\end{eqnarray}
The probability $P(S,t;\phi_{0},N)$ dose not converge to a fixed value in the limit $t \rightarrow \infty$ 
in contrast with
classical random walks. Thus we introduce time-averaged probability $\bar{P}(S;\phi_{0},N)$ defined by
\begin{eqnarray}
\bar{P}(S;\phi_{0},N) \equiv \lim_{T \rightarrow \infty} \frac{1}{T} \sum_{t=0}^{T-1} P(S,t;\phi_{0},N).
\label{eqn:avP}
\end{eqnarray}
Let us calculate $\bar{P}(S;\phi_{R},N)$, which is the time-averaged probability starting a pure state $\phi_{R}$.
Submitting the wave function with coefficients (\ref{eqn:C1})-(\ref{eqn:Rc}) into the definition (\ref{eqn:Ps}), 
we find the cross terms
in the form $\lambda_{n,m,k} \lambda_{n',m',k'}$. 
If the eigenvalue $\lambda_{n,m,k}$ is different from $\lambda_{n',m',k'}$, the time-averaged value $
\lim_{T \rightarrow \infty} \sum_{t=0}^{T-1} (\lambda_{n,m,k})^{t} (\lambda_{n',m',k'}^{\ast})^{t}/T$ vanishes. 
Thus we have
\begin{eqnarray}
\bar{P}(S;\phi_{0},N)&=&\frac{1}{N^{4}}
\left[
\left |\sum_{j=1}^{4N^{2}} C_{i,j,1}  \right |^{2}
+
\left |\sum_{j=1}^{4N^{2}} C_{i,j,-1} \right |^{2}
+
\sum_{n=1}^{\frac{N-1}{2}}
\sum_{k=3}^{4} \left| \sum_{j=1}^{4N^{2}}c_{i,j,n,0,k} \right|^{2}
\right. \nonumber \\
&&
\left.
+
\sum_{n=1}^{N-1}
\sum_{k=3}^{4} \left |\sum_{j=1}^{4N^{2}}c_{i,j,n,n,k} \right |^{2}
+
\sum_{n=1}^{\frac{N-3}{2}} 
\sum_{m=n+1}^{\frac{N-1}{2}}
\sum_{k=3}^{4} \left|\sum_{j=1}^{4N^{2}}c_{i,j,n,m,k} \right |^{2}\right]. 
\label{eqn:avPS}
\end{eqnarray}
Plugging equations (\ref{eqn:C1})-(\ref{eqn:Rc}) into (\ref{eqn:avPS}), the time-averaged probability
$\bar{P}(R;\phi_{R},N)$ is given by
\begin{eqnarray}
\bar{P}(R;\phi_{R},N)=\frac{1}{8}+\frac{5}{4N^{2}}-\frac{2}{N^{3}}+\frac{5}{4N^{4}}.
\label{eqn:avPR}
\end{eqnarray}
The time-averaged probability $\bar{P}(S;\phi_{R},N)$ is a monotone decreasing function in $N$ and it converges to 1/8 in the 
limit $N \rightarrow \infty$. Thus $\bar{P}(S;\phi_{R},N)$ is larger than 1/8 for any odd $N$. It means that the quantum walker centralizes at the origin.

We must pay attention to the dependence of the wave function on the parity of time. The value of wave function at odd time is small in comparison with the value at even time. It is similar to the fact that the probability of return to the origin at odd time is zero in a classical random walk on an infinite square
lattice. Let  $\bar{P}_{e}(S;\phi_{R},N)$ and $\bar{P}_{o}(S;\phi_{R},N)$ be the time-averaged probabilities over even time and odd time, respectively.  Then we have 
\begin{eqnarray}
\bar{P}_{e}(R;\phi_{R},N) &=&\frac{1}{4}+\frac{3}{2N^{2}}-\frac{2}{N^{3}}+\frac{5}{4N^{4}}, \\
\bar{P}_{o}(R;\phi_{R},N) &=&\frac{1}{N^{2}}-\frac{2}{N^{3}}+\frac{5}{4N^{4}}.
\label{eqn:avPReo}
\end{eqnarray}
The probability averaged over odd time $\bar{P}_{o}(R;\phi_{R},N)$ converges zero in the limit $N \rightarrow \infty$. Thus there
is a relation $\bar{P}_{e}(R;\phi_{R},N)=2 \bar{P}(R;\phi_{R},N)$ in the limit $N \rightarrow \infty$.

\subsection{The time-averaged probability in the limit $N \rightarrow \infty$}

We now proceed to the probability $P(S,t;\phi_{R},N)$
 in the limit $N \rightarrow \infty$. The only first and second terms in  
(\ref{eqn:avPS}) remain  in the limit of $N \rightarrow \infty$. 
Thus its calculation becomes easy as shown below:
\begin{eqnarray}
\bar{P}_{\infty}(S;\phi_{R}) & \equiv & \lim_{N \rightarrow \infty}
\bar{P}(S;\phi_{R},N), \\
&=&\lim_{N \rightarrow \infty}
\frac{1}{N^{4}}
\left(
\left | C_{i,0,1}  \right |^{2}
+
\left | C_{i,0,-1} \right |^{2}
\right ).
\label{eqn:avPRinf}
\end{eqnarray}
Furthermore the constant and the single summation in $C_{i,j,1}$ and $C_{i,j,-1}$  do not
contribute to $\bar{P}_{\infty}(S;\phi_{R})$. Consequently, the probability
$\bar{P}_{\infty}(S;\phi_{R})$ is given by
\begin{eqnarray}
\bar{P}_{\infty}(S;\phi_{R}) =
\left (\lim_{N \rightarrow \infty}\frac{1}{N^{2}}
\sum_{n=1}^{\frac{N-3}{2}} 
\sum_{m=n+1}^{\frac{N-1}{2}} c_{i,0,n,m,1}
\right )^{2}
+
\left(\lim_{N \rightarrow \infty}\frac{1}{N^{2}}
\sum_{n=1}^{\frac{N-3}{2}} 
\sum_{m=n+1}^{\frac{N-1}{2}} c_{i,0,n,m,2}
\right )^{2}.
\label{eqn:Pavsimp}
\end{eqnarray}

Let consider  time-averaged probabilities in $N \rightarrow \infty$ for all possible states at the origin.
The value of $c_{i,0,n,m,1}$ and $c_{i,0,n,m,2}$ as function $n$ and $m$ are given by
\begin{eqnarray}
c_{2,0,n,m,1} &=& \frac{2\,\left( \cos \xi_{m} + 
      \cos \xi_{n} - 
      2\,\cos \xi_{m}\,
       \cos \xi_{n} \right) }{-2 + 
    \cos \xi_{m} + 
    \cos \xi_{n}}, \\
c_{2,0,n,m,2} &=& \frac{2\,\left( \cos \xi_{m} + 
      \cos \xi_{n} + 
      2\,\cos \xi_{m}\,
       \cos \xi_{n} \right) }{2 + 
    \cos \xi_{m} + 
    \cos \xi_{n}}, \\
c_{i,0,n,m,1} &=& 
\frac{8 \sin^{2} (\xi_{m}/2) \sin^{2} (\xi_{n}/2)}
{2 - \cos \xi_{m} -\cos \xi_{n}}
 \hspace{3mm} \mbox{for} \hspace{3mm} i=3,4,\\
c_{i,0,n,m,2} &=& \frac{8 \cos^{2} (\xi_{m}/2) \cos^{2} (\xi_{n}/2)}
{2 + \cos \xi_{m} + \cos \xi_{n}}  \hspace{3mm} \mbox{for} \hspace{3mm} i=3,4.
\label{eqn:Gcs}
\end{eqnarray}
The double summations in (\ref{eqn:Pavsimp}) in the limit $ N \rightarrow \infty$
are calculated by replacing the summations into the following integrals:
\begin{eqnarray}
\lim_{N \rightarrow \infty}\frac{1}{N^{2}}
\sum_{n=1}^{\frac{N-3}{2}} 
\sum_{m=n+1}^{\frac{N-1}{2}} c_{2,0,n,m,1}
&=&\frac{1}{8\pi^{2}}
\int_{0}^{\pi}dx\int_{0}^{\pi}dy
\frac{2\,\left( \cos x + \cos y - 
2\,\cos x\,\cos y \right) }{-2 + \cos x + 
    \cos y}, \nonumber \\
&=& \frac{1}{4}-\frac{1}{\pi}, \\
\lim_{N \rightarrow \infty}\frac{1}{N^{2}}
\sum_{n=1}^{\frac{N-3}{2}} 
\sum_{m=n+1}^{\frac{N-1}{2}} c_{3,0,n,m,1} 
&=&\frac{1}{2\pi^{2}}
\int_{0}^{\frac{\pi}{2}}dx\int_{0}^{\frac{\pi}{2}}dy
\frac{8 \sin^{2} x \sin^{2} y}
{2 - \cos 2 x - \cos 2 y} \nonumber, \\
&=& \frac{1}{4}-\frac{1}{2\pi}.
\label{eqn:integrals1}
\end{eqnarray}
Similarly we can calculate the summation for $c_{2,0,n,m,2}$ and
$c_{3,0,n,m,2}$, and we have the same values $1/4-1/\pi$ and
$1/4-1/2\pi$. Finally the time-averaged probabilities are given by
\begin{eqnarray}
\bar{P}_{\infty}(R;\phi_{R}) &=& \frac{1}{8}, \\
\bar{P}_{\infty}(L;\phi_{R}) &=& \frac{1}{8}+\frac{2}{\pi^2}-\frac{1}{\pi},  \\
\bar{P}_{\infty}(U;\phi_{R}) &=& \frac{1}{8}+\frac{1}{2\pi^2}-\frac{1}{2\pi}, \\
\bar{P}_{\infty}(D;\phi_{R}) &=& \frac{1}{8}+\frac{1}{2\pi^2}-\frac{1}{2\pi}.  
\label{eqn:avPRs}
\end{eqnarray}
Summing $\bar{P}_{\infty}(S;\phi_{R})$  over all possible states, 
the time-averaged 
probability that a quantum walker exists at the origin is 
\begin{eqnarray}
\bar{P}_{\infty}(\phi_{R}) &=& 
\bar{P}_{\infty}(R;\phi_{R})+\bar{P}_{\infty}(L;\phi_{R})+
\bar{P}_{\infty}(U;\phi_{R})+\bar{P}_{\infty}(D;\phi_{R}), \nonumber \\
    &=& \frac{1}{2}+\frac{3}{\pi^2}-\frac{2}{\pi} 
\end{eqnarray}

\subsection{Dependence of the time-averaged probability on an initial state}

If the Grover walk starts from the pure state, then 
the localization of the quantum walker at the origin was shown 
in the previous section. The numerical results obtained by Tregenna et al. \cite{Tregenna2003}, however, suggests that
the time-averaged probability at origin of the Grover walk with a certain mixed initial state
becomes zero. To confirm this observation we consider 
the wave function at the origin starting the next mixed initial state
\begin{eqnarray}
\phi(\alpha, \beta) =\alpha |R,0,0,0\rangle + \beta |L,0,0,0\rangle,
\label{eqn:Istate}
\end{eqnarray}
where $|\alpha|^{2}+|\beta|^{2}=1$.
In the formula in (\ref{eqn:wf2}) the coefficients corresponding to $j>2$ are zero and the coefficients corresponding to $j=1$ are already computed in (\ref{eqn:Rc}). Thus we consider 
the only coefficients corresponding to $j=2$.  Suppose that $\psi(0)=\phi_{L}=
(0,1,0,\ldots,0 )^{T}$. Then we obtain
\begin{eqnarray}
C_{1,2,1}&=& -\frac{7}{4}+\frac{3N}{2}
-\sum_{n=1}^{\frac{N-1}{2}} \frac{8}{3+\cos\xi_{n}} \nonumber \\
& & \hspace{10mm}
+\sum_{n=1}^{\frac{N-3}{2}}\sum_{m=n+1}^{\frac{N-1}{2}}
\frac{2\,\left( \cos \xi_{m} + \cos \xi_{n} + 2\,\cos \xi_{m}\,\cos \xi_{n} \right) }
{2 + \cos \xi_{m} + \cos \xi_{n}}, 
\label{eqn:C2} \\
C_{1,2,-1}&=& \frac{3}{4}-\frac{N}{2}
+\sum_{n=1}^{\frac{N-3}{2}}\sum_{m=n+1}^{\frac{N-1}{2}}
\frac{2\,\left( \cos \xi_{m} + \cos \xi_{n} - 2\,\cos \xi_{m}\,\cos \xi_{n} \right) }
{-2 + \cos \xi_{m} + \cos \xi_{n}}, \\
c_{1,2,n,0,3}&=& c_{1,2,n,0,4}=-1+\frac{4}{3+\cos{\xi_{n}}}, \\
c_{1,2,n,n,3}&=& c_{1,2,n,n,4}=0, \\
c_{1,2,n,m,3}&=& c_{1,2,n,m,4}=\frac{4(\cos\xi_{m}-\cos\xi_{n})^{2}}
{6-\cos (2\xi_{m})-\cos (2\xi_{n})-4\cos\xi_{n}\cos\xi_{m}} \hspace{3mm} \mbox{for}
\hspace{3mm} m>n.
\label{eqn:Gc}
\end{eqnarray}
We now can calculate the time-averaged probability $\bar{P}(R;\phi(\alpha, \beta),N)$ for any odd system size $N$ by combining (\ref{eqn:C1}) - (\ref{eqn:Rc}) with (\ref{eqn:C2}) - (\ref{eqn:Gc}). However it is rather complicated. 
Thus we show only the result in the limit of $N \rightarrow \infty$.
The time-averaged probability $\bar{P}_{\infty}(R;\phi(\alpha,\beta))$  is obtained by replacing the summations to the integrals in the same way
described in the previous section, and it becomes
\begin{eqnarray}
\bar{P}_{\infty}(R;\phi(\alpha,\beta))
=\frac{1}{8}\left| \alpha+ \left(1-\frac{4}{\pi}\right)  \beta \right|^{2}.
\label{eqn:Pavgen1}
\end{eqnarray}
Similarly we have the time-averaged probability corresponding to the ``L" state
\begin{eqnarray}
\bar{P}_{\infty}(L;\phi(\alpha,\beta))
=\frac{1}{8}\left| \beta+ \left(1-\frac{4}{\pi}\right)  \alpha \right|^{2}.
\label{eqn:Pavgen2}
\end{eqnarray}
Fig. 2 shows  $\bar{P}_{R}(\alpha) \equiv \bar{P}_{\infty}(R;\phi(\alpha,\sqrt{1-\alpha^2}))$  and 
$\bar{P}_{L}(\alpha) \equiv \bar{P}_{\infty}(L;\phi(\alpha,\sqrt{1-\alpha^2}))$
for $\alpha \in [-1,1]$. The time-averaged probability $\bar{P}_{R}(\alpha)$ becomes zero at $\alpha_{min}$ given by
\begin{eqnarray}
\alpha_{min} &=& \sqrt{1-\frac{\pi^2}{16-8\pi+2\pi^2}}, 
\end{eqnarray}
and it takes the maximum value
at $\alpha_{max}$ given by
\begin{eqnarray} 
\alpha_{max} &=& \frac{\pi}{\sqrt{16-8\pi+2\pi^2}}.
\end{eqnarray}
Since the value $\alpha=1$ and $\alpha=-1$ indicate the pure state, 
it is found that the time-averaged probability $\bar{P}_{R}(\alpha)$
takes the maximum at the mixed initial state.  As shown in the numerical calculation
\cite{Tregenna2003}, the quantum walk starting from a certain
initial state spreads out.  Although the time-averaged probability $\bar{P}_{R}(\alpha)$ becomes zero at 
$\alpha=\alpha_{min}$, the quantum walker remains at the origin. Put another way, the component of the time-averaged probability
corresponding to the ``R" state converges to zero,
 but other component remains positive.  Therefore we can say that the quantum walker in the state ``R" perfectly
converts into other state in the limit $t \rightarrow \infty$ at $\alpha=\alpha_{min}$. 

Let us consider the time-averaged probability for more general initial states defined by
\begin{eqnarray}
\phi(\alpha, \beta, \gamma, \zeta) =\alpha |R,0,0,0\rangle + \beta |L,0,0,0\rangle + \gamma |U,0,0,0\rangle + 
\zeta |D,0,0,0\rangle,
\label{eqn;Istate}
\end{eqnarray}
where $|\alpha|^{2}+|\beta|^{2}+|\gamma|^{2}+|\zeta|^{2}=1$.
Repeating the procedure described above yields  the time-averaged probability for this initial state:  
\begin{eqnarray}
\bar{P}_{\infty}(R;\phi(\alpha,\beta,\gamma,\zeta))=
\left| \frac{\alpha}{2\,{\sqrt{2}}} - {\sqrt{\frac{1}{8} + \frac{2}{{\pi }^2} - \frac{1}{\pi }}}\beta\ 
+ {\sqrt{\frac{1}{8} + \frac{1}{2\,{\pi }^2} - \frac{1}{2\,\pi }}}\,(\gamma +\zeta)
\right|^{2}.
\label{eqn:PavFullg}
\end{eqnarray}

Taking symmetry into account, the other components are given by
\begin{eqnarray}
\bar{P}_{\infty}(L;\phi(\alpha,\beta,\gamma,\zeta)) &=& \bar{P}_{\infty}(R;\phi(\beta,\alpha,\gamma,\zeta)), \\
\bar{P}_{\infty}(U;\phi(\alpha,\beta,\gamma,\zeta)) &=& \bar{P}_{\infty}(R;\phi(\gamma,\zeta,\alpha,\beta)), \\
\bar{P}_{\infty}(D;\phi(\alpha,\beta,\gamma,\zeta)) &=& \bar{P}_{\infty}(R;\phi(\zeta,\gamma,\alpha,\beta)).
\label{eqn:PavFullg2}
\end{eqnarray}

The condition for which all components in (\ref{eqn:PavFullg}) - (\ref{eqn:PavFullg2}) become zero is obtained by 
\begin{eqnarray}
\alpha=e^{i \theta}/2, \hspace{3mm} \beta=\alpha, \hspace{3mm}  \gamma=-\alpha, \hspace{3mm} \zeta=-\alpha.
\label{eqn:Czero}
\end{eqnarray}
Setting $\theta=0$, we confirm the numerical results given by Tregenna et al. \cite{Tregenna2003}, which claim that the probability of $\bar{P}_{\infty}(S;\phi(\alpha,\beta,\gamma,\zeta))$ becomes zero
for any state $S$ by setting $\alpha=\beta=-\gamma=-\zeta=1/2$.
Fig. 3 shows a snapshot after $30$ steps on  a square lattice with $N=51$ starting from
an initial state with $\theta=1/3$ in (\ref{eqn:Czero}). One definitely finds that the localization
disappears. On the other hand, the summation of 
$\bar{P}_{\infty}(S;\phi(\alpha,\beta,\gamma,\zeta))$ over all possible states takes a maximum
values $2+8/\pi^2-8/\pi$=0.26409.. at $\alpha=\beta=\gamma=\zeta=1/2$. Thus  the time-averaged probability over even time is larger than 1/2.

\subsection{Conditions of the localization}

We saw that a spike exists at the origin in Grover walk, but the two-dimensional quantum walkers
governed by the matrix $A_{1}$ and $A_{2}$ spread out. The significant difference between the Grover walk
and  other quantum walks is the degree of eigenvalues. For example, the eigenvalues of $H_{n,m}$
with $A_{1}$ are given by
\begin{eqnarray}
\left[
\begin{array}{r}
\lambda_{n,m,1} \\
\lambda_{n,m,2} \\
\lambda_{n,m,3} \\
\lambda_{n,m,4}
\end{array} 
\right]
=
\left[
\begin{array}{r}
 \sqrt{i\cos\xi_{n} \cos\xi_{m}+f_{n,m}},  \\
-\sqrt{i\cos\xi_{n} \cos\xi_{m}+f_{n,m}},  \\
\sqrt{i\cos\xi_{n} \cos\xi_{m}-f_{n,m}},  \\
-\sqrt{i\cos\xi_{n} \cos\xi_{m}-f_{n,m}},  \\
\end{array}
\right],
\end{eqnarray}
where
\begin{eqnarray}
f_{n,m}=\sqrt{\sin^{2}\xi_{m}+\cos^{2}\xi_{n}\cos^{2}\xi_{m}}.
\end{eqnarray}
We find no common eigenvalues to all value $n$ and $m$ 
such as $-1$ and $1$ in $\lambda_{n,m,k}$. If a quantum walker exists only at the origin initially, 
the coefficient $c_{i,j,n,m,k}$ 
in  (\ref{eqn:avPS}) takes non-zero value
for only one  $j$.  Therefore the summation over $j$ in  (\ref{eqn:avPS}) does not depend on the system size and
the summation over $n$ and $k$  increases  with the system size. 
Accordingly, the order of third and fourth term in  (\ref{eqn:avPS}) is $\cal O \mit (N^{-3})$. 
Similarly the order of the
fifth term in  (\ref{eqn:avPS}) is $\cal O \mit (N^{-2})$.
 As a result, these terms vanish in the limit $N \rightarrow \infty$.
As shown in (\ref{eqn:Cp}) and (\ref{eqn:Cm}), since $C_{i,j,1}$ and $C_{i,j,-1}$ contain double summations, 
the order of them is $\cal O \mit (N^{2})$. This large contribution to the time-averaged probability
comes from the fact that the eigenvalues $-1$ and $1$ exist for any $n$ and $m$ in common. 
From the above consideration one conjectures that there is another two-dimensional quantum walk in which the walker centralize at the origin. Thus we find a new matrix $A_{4}$ 
such that $H_{n,m}(A_{4})$ contains
eigenvalue $-1$ and $1$ independently on the value $n$ and $m$. Suppose that the matrix $A_{4}$ is real symmetric matrix. Then a possible matrix is given by
\begin{eqnarray}
A_{4}=
\left[
\begin{array}{cccc}
-p             & q             &  \sqrt{pq}  & \sqrt{pq} \\
q            &  -p             &  \sqrt{pq}  & \sqrt{pq} \\
\sqrt{pq}    &  \sqrt{pq}    &  -q           & p         \\
\sqrt{pq}    &  \sqrt{pq}    &  p          &  -q
\end{array}
\right],
\label{eqn:new}
\end{eqnarray}
where $q=1-p$. The Grover walk is corresponding to $p=1/2$.
As an example we show the probability $P(x,y,t;\phi_{R})$ of a quantum walk 
which is characterized by the matrix $A_{4}$ with
$p=1/3$ and $q=2/3$ in  Fig.4.  We clearly recognize the existence of a spike at the origin.

\section{Summary}
We have shown analytically that the localization of the Grover walk at the initial position
can be surely measured for any odd system size.  On the other hand,  we have also shown that
we can chose special initial states with which the quantum walk disappears at the initial position
in $N \rightarrow \infty$. As pointed out by Tregenna et al. \cite{Tregenna2003}, 
this different behavior can be used to control
the Grover's search. We here summarize the reason why the localization exists in the Grover walk 
starting  from a local position.  

There are
$16N^{2}$ quantum states in the Grover walk on the square lattice including $N^{2}$ sites. Therefore the number
of eigenvalues and eigenvectors of the time-evolution operator is also $16N^{2}$. Since the wave function at time $t$
is expressed by a liner combination of the $t$-th power of the eigenvalue in which the coefficients
are given by the product of component of eigenvectors. Each component of the eigenvector decrease in inverse ratio to the
system size and the coefficients decrease in proportion to $N^{-2}$.
Thus, if the quantum walk exists initially at an fixed point and all eigenvalues are distinct, then the probability 
of observing the quantum walk at the fixed point goes to zero by taking the system size infinity.
For this reason,  the degeneration of eigenvalues is necessary for the localization.  
In the case of 
quantum walks on a circle including odd sites, 
the eigenvalues are distinct except for trivial cases \cite{Aharonov1998}. Thus the localization is not observed.

The eigenvalues of the Grover walk include $-1$ and $1$, and the degree of the degeneration of them are $N^{2}+2$
and $N^{2}$, respectively.  As mentioned above, the each coefficient itself decrease in the form $N^{-2}$, but
the  degree of the degeneration is proportion $N^{2}$. As a consequence, the eigenvalue $-1$ and $1$
can positively contribute to the probability even if $N \rightarrow \infty$ except 
for the case that both coefficients of $-1$ and $1$ become zero. 
If we choose such initial states as   both coefficients of $-1$ and $1$ corresponding to the
wave function at the origin are zero for all states, then the quantum walker spreads
from the origin.

\begin{acknowledgments}
The authors wish to thank Hiroshi Araki for simulations.
\end{acknowledgments}

\clearpage

\begin{figure}
\includegraphics{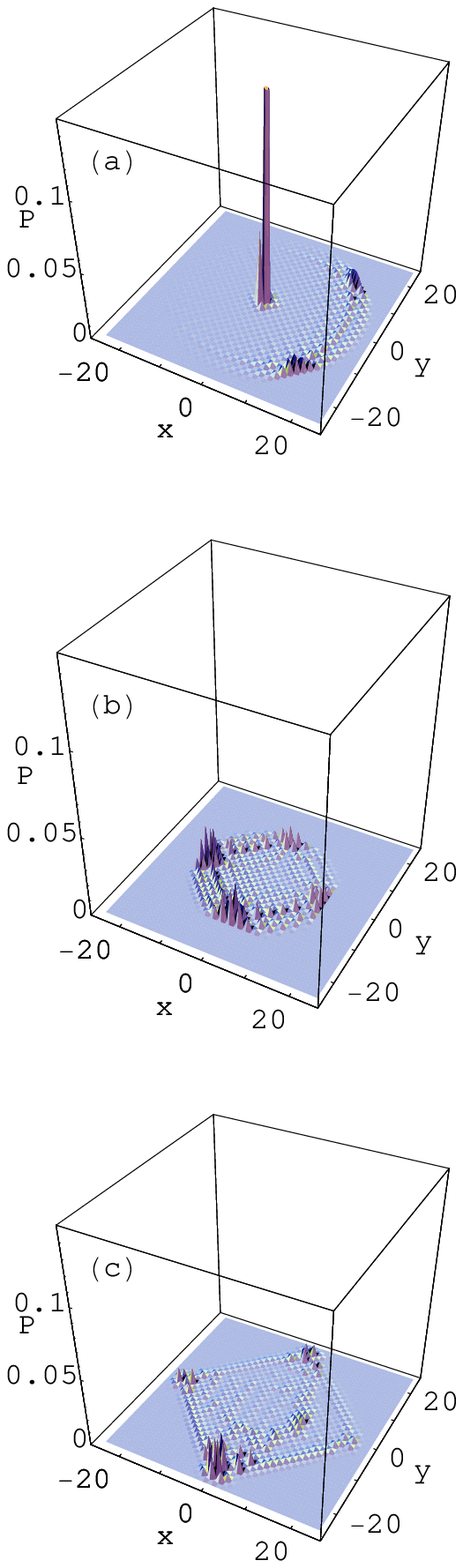}
\caption{\label{fig:Figure1} 
Snapshots of probability distribution $P(x,y,t;\Psi(0))$ at time $t=30$ 
on a sure lattice with system size $N=51$ with $\Psi(0)=(1,0,\ldots,0)^{T}$.
 Each of evolutions are determined by the matrices
 $A_{0}$ (Grover walk),
$A_{1}$ and $A_{2}$ defined in (\ref{eqn:matA0})-(\ref{eqn:matA1}) 
 corresponding to (a), (b) and (c). The central peak in (a) shows the localization of
the Grover walk.
}
\end{figure}

\clearpage

\begin{figure}
\includegraphics{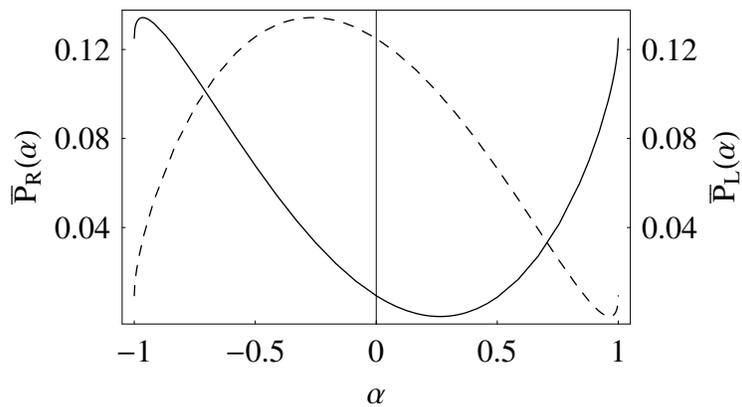}
\caption{\label{fig:Figure2} 
Solid line denotes the time-averaged probability $\bar{P}_{R}(\alpha)$ and 
dashing line denotes $\bar{P}_{L}(\alpha)$. The value of 
$\bar{P}_{R}(\alpha)$ at $\alpha=1$ and 1 is $1/8$, and 
$\bar{P}_{R}(\alpha)$ takes zero at $\alpha_{min}=0.26357..$.
}
\end{figure}

\clearpage

\begin{figure}
\includegraphics{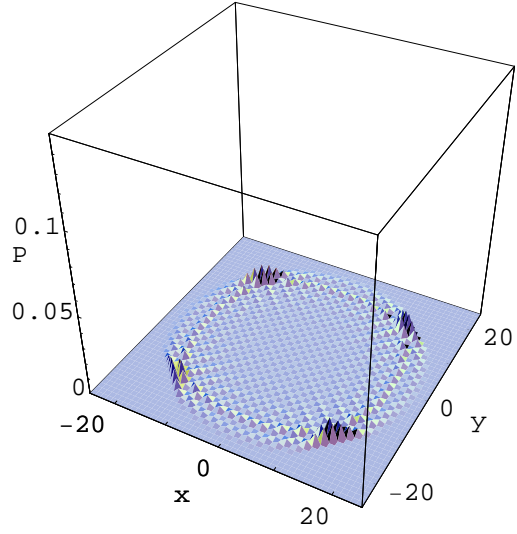}
\caption{\label{fig:Figure3} 
The probability $P(x,y,t;\Psi(0))$ of Grover walk at $t=30$ starting a mixed state
$\Psi(0)=( e^{i/3}, e^{i/3}, -e^{i/3},-e^{i/3},0, \ldots, 0 )^{T}$ with $N=51$. 
 The central localization in Fig. 1 (a) disappears.
}
\end{figure}

\clearpage

\begin{figure}
\includegraphics{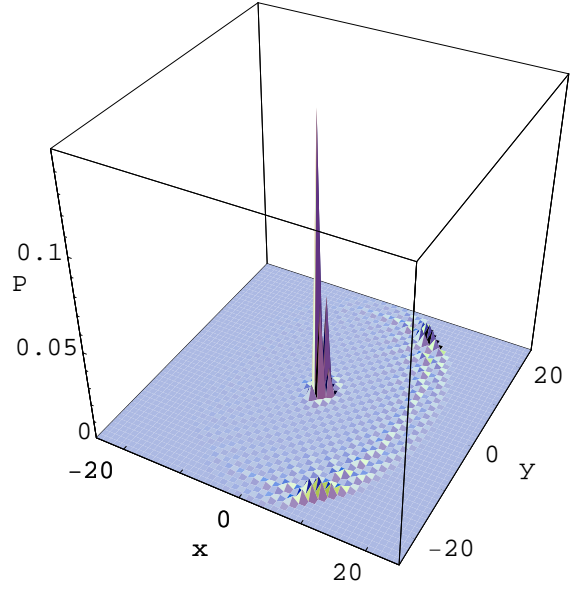}
\caption{\label{fig:Figure4} 
The probability $P(x,y,t;\phi_{R})$ of the quantum walk whose wave function is determined
by the matrix $A_{4}$ with $p=1/3$ and $q=2/3$.
 A sharp distribution at the origin is observed similarly as in the case of the Grover walk.
}
\end{figure}

\end{document}